\documentclass[preprint,NumberedRefs]{JASA}

\begin{document}

\title{Rapid Method for Computing the Mechanical Resonances of Irregular Objects}

\author{Avi Shragai}
\affiliation{Laboratory of Atomic and Solid State Physics, Cornell University, Ithaca, NY 14853, USA}
\author{Florian Theuss}%
\affiliation{Laboratory of Atomic and Solid State Physics, Cornell University, Ithaca, NY 14853, USA}
\author{Ga\"el Grissonnanche}
\affiliation{Laboratory of Atomic and Solid State Physics, Cornell University, Ithaca, NY 14853, USA}
\affiliation{Kavli Institute at Cornell for Nanoscale Science, Ithaca, NY 14853, USA}
\author{B. J. Ramshaw}
\email{bradramshaw@cornell.edu}
\affiliation{Laboratory of Atomic and Solid State Physics, Cornell University, Ithaca, NY 14853, USA}

\date{\today}

\begin{abstract}
\nolinenumbers

    A solid object’s geometry, density, and elastic moduli completely determine its spectrum of normal modes. 
    Solving the inverse problem -- determining a material's elastic moduli given a set of resonance frequencies 
    and sample geometry -- relies on the ability to compute resonance spectra accurately and efficiently. 
    Established methods for calculating these spectra are either fast but limited to simple geometries, or are
    applicable to arbitrarily shaped samples at the cost of being prohibitively slow. Here, we describe a 
    method to rapidly compute the normal modes of irregularly shaped objects using entirely 
    open-source software. Our method's accuracy compares favorably with existing methods for simple geometries and 
    shows a significant improvement in speed over existing methods for irregular geometries. 
     
\end{abstract}

%% pacs numbers not used

\maketitle
\nolinenumbers

%  End of title page for Preprint option --------------------------------- %

\section{\label{sec:1} Introduction}

Accurate measurements of elastic moduli are required for both fundamental 
research and for engineering applications. The behavior of elastic moduli as a function of temperature or 
magnetic field has been used to study unconventional superconductors, heavy fermion metals, quantum magnets, and other correlated states of matter \cite{sr2ruo4_sc, ybco_pseudogap, pucoga5_valence, mn3x_strong, cu_glass, op_coupling_alloys,layered_perovskite,doubled_perovskite,skyrmion_mnsi,skyrmion_creep}. Accurate measurements of elastic moduli also allow for precision quality control in manufactured parts, for example in ball bearings used in aeronautics, or monitoring for damage in large-scale steel parts \cite{rus_bearings, rus_complex, rus_testing, rus_bridge}.

Resonant ultrasound spectroscopy (RUS) is a non-destructive method for measuring all components of a 
material's elastic tensor simultaneously \cite{rus_toolbox, modern_rus}. In an RUS experiment, a sample of known crystalline symmetry is 
driven with small-amplitude, oscillatory stresses under free boundary conditions, exciting its mechanical resonances. The elastic moduli are determined by fitting the measured resonances to calculated spectra. 

For samples with simple geometric shapes, the Rayleigh-Ritz method developed by Visscher \textit{et al.} \citet{visscher_rus} allows rapid and accurate computation 
of resonance spectra---the forward computation---with relatively 
modest computational resources. If a good initial guess for the elastic moduli is available, then a fit of the 
elastic constants can take as little as a few minutes. However, when a sample cannot be polished into 
a rectangular prism, cylinder, or sphere, other, more computationally expensive methods such as finite elements 
must be used. For irregularly shaped samples with low crystal symmetry, a single forward computation using finite elements can 
take tens of seconds. Then, if no good initial guesses for the elastic moduli are available, a heuristic fitting algorithm such as a genetic algorithm is required \cite{pucoga5_valence} and the inversion procedure may require thousands of forward computations. A fit can then take tens of hours if finite elements are used for the forward computation \cite{fem_arbitrary_rus,fem_fpi_rus,fem_dif_evolution_rus}.

It is therefore crucial to develop alternative methods for computing the mechanical resonances of irregularly 
shaped objects. Here, we outline a method---referred to hereafter as the ``surface-mesh integration'' (SMI) method---that generalizes the approach developed by Visscher \textit{et al.}\citet{visscher_rus} for performing the forward calculation for samples with irregular geometry. We assume that the surface of the sample has been digitized, for example by micro-computed tomography \cite{fem_muct_rus}. SMI allows all of the computationally intensive steps of the forward calculation to be done once, after which the elastic moduli can be varied and the resonance frequencies calculated rapidly until a fit is achieved, as in the Visscher method. We compare the SMI method to both the Visscher and finite element methods and find that the results are identical to within experimental precision, with the SMI method being several orders of magnitude faster than finite elements.

\section{\label{sec:2} Method}

\begin{figure}[t]
    \centering
    \includegraphics[width=.5\textwidth]{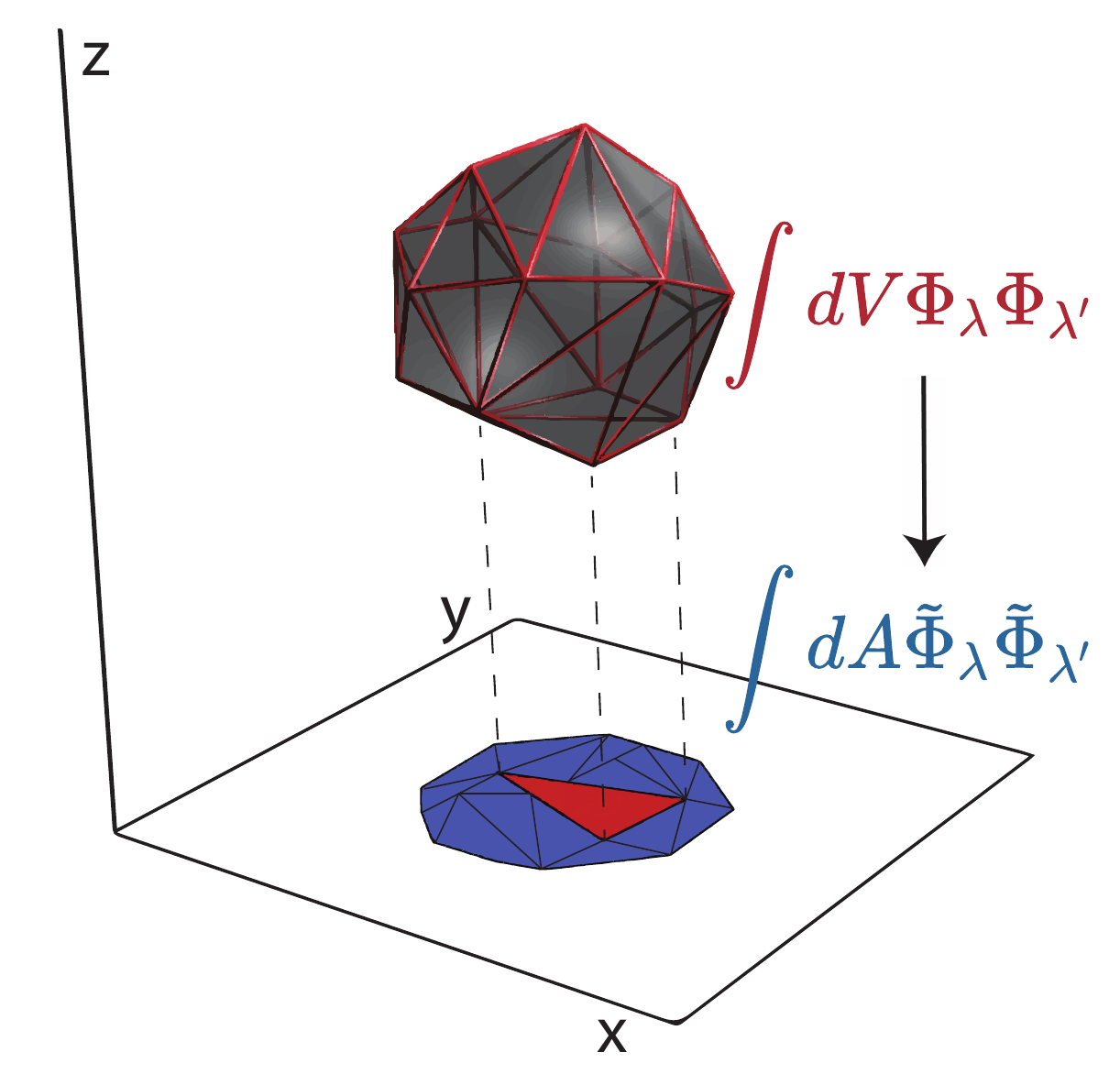}
    \caption{\textbf{Visualizing surface-mesh integration.} Both SMI and Visscher's method for computing mechanical resonances require tabulating integrals of polynomials over the interior of an object. To accomplish this numerically, we create a triangular mesh of the object's boundary and perform the equivalent surface integration. As described in \protect\citet{bem_wu}, each mesh element is projected onto a plane and mapped to a reference triangle, after which the integral is computed using quadrature rules. For example, the integral of the polynomial $\Phi_\lambda\Phi_{\lambda'}$ over the volume of the coarse red wireframe is transformed to an integral of the function $\tilde{\Phi}_\lambda\tilde{\Phi}_{\lambda'}$ over the wireframe's projection onto the x-y plane, shown in blue, using Gauss's theorem. The integral is evaluated as a sum of contributions from triangular mesh elements. We illustrate one of these mesh elements and its projection in the x-y plane by connecting their vertices with dotted lines in the schematic above.}
    \label{fig:stokes_schematic}
\end{figure}

The Lagrangian for an object with uniform density $\rho$ and elastic tensor $C_{ijkl}$ is written in  terms of its harmonic displacement $u$ from equilibrium at frequency $\omega$ as
\begin{equation}
\label{eqn:wave_lagrangian}
L = \int dV\left(\frac{1}{2} \rho \omega^2 u_i u_i - \frac{1}{2}\frac{\partial u_{i}}{\partial x_j}C_{ijkl}\frac{\partial u_{l}}{\partial x_k}\right),
\end{equation}
where repeated indices are summed over the three spatial coordinates $x$, $y$, and $z$, and integration is performed over the sample volume. Following Visscher \textit{et al.} 
\citet{visscher_rus}, the components of the displacement are expanded in a polynomial basis. This basis is commonly a simple power basis with basis functions $\Phi_{\lambda} \equiv x^n y^m z^{\ell}$, where $\lambda$ labels the particular combination of $n$, $m$, and $\ell$. The components of the displacement are then given by a linear combination of basis functions: $u_i = a_{i\lambda}\Phi_{\lambda}$. 

As with other variational methods for computing mechanical resonances, including finite elements, the Visscher method converges to the exact eigenfrequencies, $\omega$, from \autoref{eqn:wave_lagrangian} as more basis functions are added \cite{visscher_rus}. In practice, the basis is cut off at order $N \geq n+m+\ell$, where $N$ is high enough to ensure that the resonance frequencies are computed to better than the experimental uncertainty due, for example, to uncertainty in the sample dimensions. Typically, these uncertainties are approximately one-tenth of a percent and require $N\approx 15$ or better to fit the first $\sim\!100$ resonances. \cite{sr2ruo4_sc,mn3x_strong}

Inserted into the Euler-Lagrange equations, \autoref{eqn:wave_lagrangian} gives a generalized eigenvalue problem for the resonance frequencies $\omega$ \cite{demarest_rus}:
\begin{equation}
\omega^2\rho a_{i\lambda}\delta_{il} \!\!\int\!\! dV\left(\Phi_{\lambda} \Phi_{\lambda'}\right) = a_{i\lambda}C_{ijkl} \!\!\int\!\! dV \left(\frac{\partial \Phi_{\lambda}}{\partial x_j}\frac{\partial \Phi_{\lambda'}}{\partial x_k}\right),
\label{eqn:eigval}
\end{equation}
where the left-hand side captures the kinetic energy and the right-hand side the potential energy due to 
the material stiffness. For anisotropic materials, including single crystals, the coordinate system used to perform the integration must be referenced to the coordinate axes used to specify the material's elastic tensor, $C_{ijkl}$. This relation for the resonance frequencies is exact in the absence of acoustic loss, with corrections from attenuation scale as $1/Q^2$ where $Q$ is the resonance quality factor. \cite{rus_toolbox} In high-quality samples, $Q$ is typically of order $10^4$ and these corrections are negligible compared to systematic uncertainties. \cite{rus_toolbox,sr2ruo4_sc,mn3x_strong} By indexing the basis functions $\Phi_\lambda$ (for an explicit example see \citet{rus_toolbox}), \autoref{eqn:eigval} can be assembled into a matrix equation and solved using a numerical linear algebra library. The size 
of the matrices scale as $N^3$ and the spectrum can be computed in a fraction of a 
second for $N\sim 15$. Computing resonance modes in this way reduces the problem to tabulating integrals of the basis functions and their derivatives over the sample volume. The strength of the power basis is that, for simple sample geometries, these integrals can be computed exactly. 

For irregularly shaped samples, however, the integrals in \autoref{eqn:eigval} must be evaluated numerically. The most natural first step is to partition the irregular object into tetrahedra---a procedure that is both computationally cheap and applicable to arbitrary geometries. Although volume integrals over tetrahedral elements can be performed numerically in principle, evaluation of these integrals to sufficient precision is difficult. For example, computing these integrals becomes prohibitively slow using a recursive, closed-form solution \cite{tetrahedron_int}. Alternatively, using quadrature integration limits the method to a power basis of order $N\le 10$, as quadrature rules over tetrahedra have only been computed to order 20 and the forward calculation using a polynomial basis of order $N$ requires tabulating integrals of products of two basis functions, which are polynomials up to order $2N$ \cite{tetrahedron_quad}.

\begin{figure*}[t]
    \centering
    \includegraphics[width=1.\textwidth]{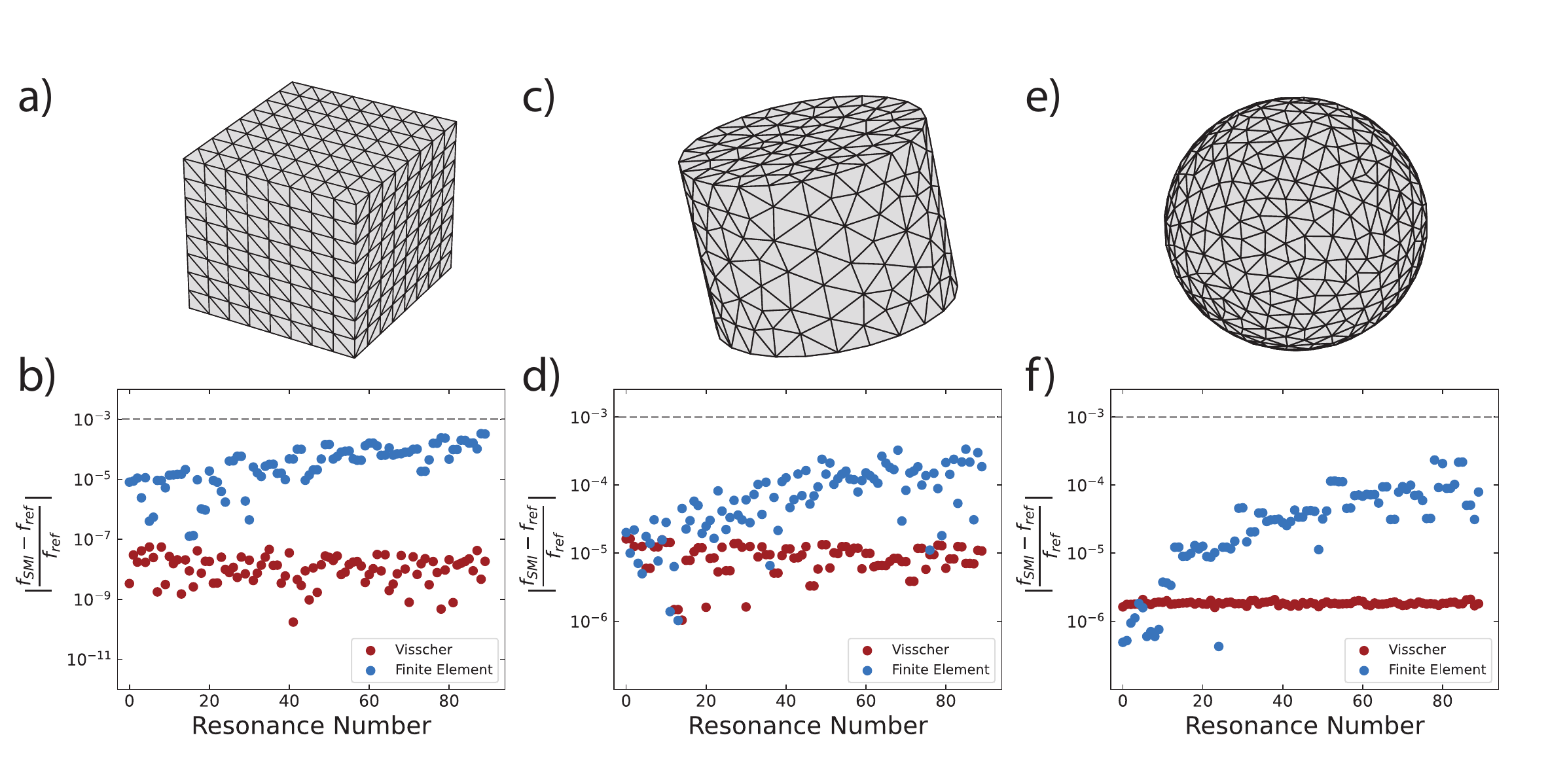}
    \caption{\textbf{Comparison between different methods for computing resonance spectra for simple geometries.} Comparisons of the first 100 resonant frequencies for three simple geometries computed by SMI ($f_{\text{SMI}}$) and the Visscher method and finite-elements. The meshed objects are shown in panels (a), (c), and (e). Below the corresponding object, in panels (b), (d), and (f), we plot $|f_{\text{SMI}} - f_{\text{ref}}|/f_{\text{ref}}$, where $f_{\text{ref}}$ are the mechanical resonances found by the Visscher and finite-element methods. There is excellent quantitative agreement between methods, and the frequency differences---a few hundredths of a percent or $\sim$ 1kHz for typical sample sizes---are nearly ten times smaller than the typical systematic uncertainties encountered in an RUS experiment, about one part-per-thousand, marked by the dashed grey lines.}
    \label{fig:compare_res}
\end{figure*}

Our new approach is to numerically evaluate the integrals in (\autoref{eqn:eigval}) by meshing only the surface of the sample, with triangular mesh elements, and rewriting  the volume integrals as integrals over the sample surface using Gauss' theorem. Our method is reminiscent of, but distinct from, boundary element methods, which map differential equations over a bounded volume onto its surface \cite{bem_theory,bem_acoustics,bem_wu,bem_anisotropy}. A triangular surface mesh allows the use of quadrature rules 
that extend to order 50---well beyond what is necessary to fit typical RUS spectra. In particular, to evaluate the integral of the product of two functions in the power basis, $\Phi_\lambda \Phi_{\lambda'} = x^n y^mz^{\ell}$, we first note that a scalar function can be written as the divergence of a vector function:
\begin{equation}
    \int dV \Phi_\lambda \Phi_{\lambda'} = \frac{1}{\ell+1}\int\! dV ~\nabla\!\cdot\! \left(0, 0, x^n y^mz^{\ell+1}\right).
\label{eqn:gauss1}
\end{equation}
Then, using the divergence theorem (``Gauss' law''), we convert the volume integral to a surface integral:
\begin{align}
  \int dV \Phi_\lambda \Phi_{\lambda'} =&\frac{1}{\ell+1}\int d\mathbf{A}\cdot \left(0, 0, x^n y^mz^{\ell+1}\right) \\
		= &\frac{1}{\ell+1}\sum_{i}\int dA_{z,i}~ x^n y^mz^{\ell+1} \\
        \equiv &\frac{1}{\ell+1}\sum_{i}\int dA_{z,i}~ \tilde{\Phi}_\lambda\tilde{\Phi}_{\lambda'},
\label{eqn:surface_int}
\end{align}
where $dA_{z,i}$ is the $i^{\rm th}$ triangular mesh element projected onto the $x-y$ plane (see \autoref{fig:stokes_schematic}). Note that the choice of plane to project onto is arbitrary and we find no significant changes in the computed resonant frequencies when choosing different projection planes. The integrals needed to assemble the wave equation (i.e. to construct \autoref{eqn:eigval}) can be evaluated once, in parallel for each mesh element if desired, and the results saved for use in a fit of the elastic moduli.

\section{\label{sec:3} Results and Discussion}

To validate our method of computing mechanical resonances using surface-mesh integration, we compare spectra obtained with the procedure outlined in the previous section to those calculated using the Visscher and finite element methods. In these tests, we consider millimeter-scale objects with elastic moduli of order 100 GPa, similar to the conditions encountered in typical RUS experiments on single crystals.

Comparison of surface-mesh integration to the Visscher method is only possible for relatively simple geometries where the volume integrals in the power basis can be computed exactly. We choose the most basic geometries for our comparison: rectangular prism, sphere, and cylinder. For both the Visscher and surface-mesh integration methods, we use an order 15 polynomial basis. We also compare surface-mesh integration to the finite-elements method implemented in Comsol. For the finite-elements method, we use a tetrahedral mesh with a maximum element size approximately 100 times smaller than the object dimensions. Although an even smaller mesh size will improve accuracy, the computation time quickly increases beyond what is practical for fitting elastic moduli, which requires repeated computations of the resonant frequencies. One advantage of our SMI method is that, because tabulating integrals is done only once, the surface mesh size can be made significantly smaller than the FEM counterpart without slowing down an RUS fitting procedure. 

The results of our comparisons for simple objects are plotted in \autoref{fig:compare_res}, which shows the difference between resonant frequencies calculated for the first 100 normal modes of each test geometry using surface-mesh integration, the Visscher method, and finite element methods. In all cases, the resonant frequencies match to several parts in $10^4$---well within experimental uncertainty.

\begin{figure*}[t]
    \centering
    \includegraphics[width=1.\textwidth]{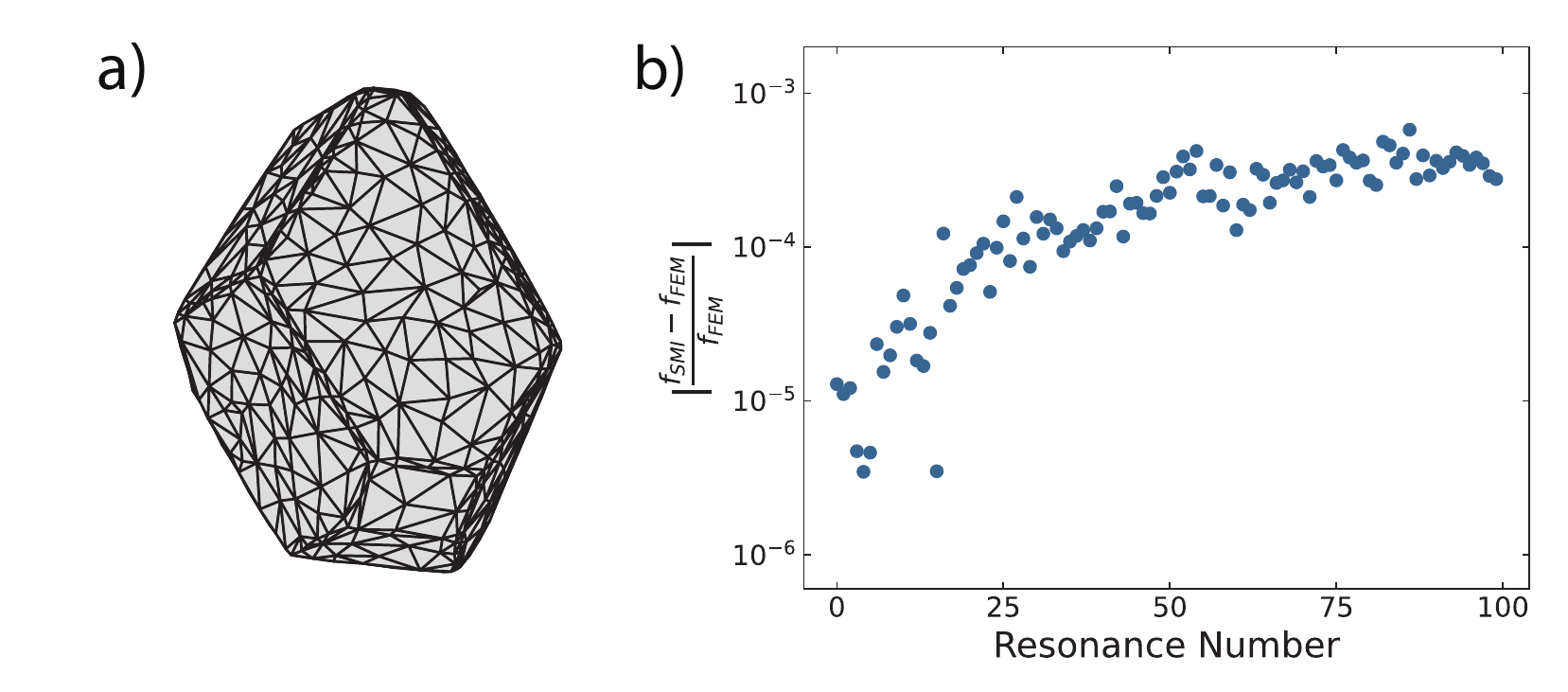}
    \caption{\textbf{Comparison of surface-mesh integration to finite elements for an irregularly shaped object.} (a) Triangular surface mesh of an irregularly shaped object. (b) Comparison of the first 100 resonant frequencies of the irregularly shaped object depicted in (a) computed using finite elements and surface mesh integration. In this case, there is no practical means of writing analytical expressions for the integrals needed to implement the Visscher method. The surface mesh integration matches the finite element result to within several hundredths of a percent using orders of magnitude less computation time.}
    \label{fig:compare_irreg}
\end{figure*}

The Visscher method is known to be accurate to better than part per million compared to closed-form solutions for an isotropic sphere and parallelepiped once the polynomial basis extends past $N\sim 10$ \cite{fem_arbitrary_rus}. This makes comparison to the Visscher method an excellent starting point for our tests. We see that the Visscher method and surface-mesh integration give nearly identical resonance frequencies across the computed spectra, with a constant offset of $\approx 10^{-5}$ for samples with curved boundaries like the sphere and cylinder. This offset is well accounted for by the volume mismatch which necessarily accompanies meshing the sphere or cylinder. The experimental uncertainties encountered when measuring rectangular prisms are typically of order $(f_{measured}-f_{calculated})/f_{measured}\approx 10^{-3}$ \cite{pucoga5_valence,sr2ruo4_sc,mn3x_strong}, suggesting that both methods are equally good for computing the spectra of simple geometric shapes.

The spectrum computed using finite elements will also converge to the exact result as the mesh size is taken to zero. In our tests, we see that finite elements and surface-mesh integration agree to better than the precision of an RUS experiment in all test geometries. In every test case, the agreement is best for the lowest frequency modes after which the resonance frequencies computed by finite elements become progressively higher than those found with the Visscher and surface-mesh integration methods. Because all of these methods approximate the modes shapes with finite degree or piecewise polynomials, a variational argument \cite{visscher_rus} suggests that over-estimation of the higher modes indicates the finite element method is losing accuracy. 

The aforementioned tests demonstrate that the surface-mesh integration method compares well with existing methods for simple objects. However, we are primarily concerned with irregularly-shaped objects, where it is impractical to write closed form expressions for the integrals needed to apply the Visscher method. We can, however, compare the surface-mesh integration method on irregular objects to the finite-elements method. An example of such an irregular object is shown in \autoref{fig:compare_irreg}a. The results of the comparison are shown in \autoref{fig:compare_irreg}b. In this case we use a polynomial basis of order 18 in the SMI calculation. Again, we see the agreement between the surface-mesh integration and finite element calculations is satisfactory for comparison with an RUS experiment. However, the distinct advantage of the surface-mesh technique is that, rather than taking several minutes, our method requires a fraction of a second to compute the spectrum of normal modes once the matrices in (\ref{eqn:eigval}) have been constructed. For example, running Comsol on an Intel$^{\circledR}$ Xenon$^{\circledR}$ Gold 2.10GHz processor requires about 50 seconds to compute the first 100 normal modes, whereas solving the generalized eigenvalue problem used in the SMI method takes only 0.5 seconds using standard linear algebra packages on the same machine. If it is necessary to compute the spectrum repeatedly, as in the RUS fitting procedure, the one-time cost of numerical integration saves hours of computation compared with finite elements---which must be re-evaluated at each step in the fit---with no loss of accuracy.

Though surface-mesh integration performs well in the test cases we describe here, there are particular geometries where finite elements will outperform surface-mesh integration. Examples include objects with small features, and objects with large voids (in both cases, ``small'' and ``large'' are in comparison to the overall size of the object). In these cases, the resonance frequencies computed by surface-mesh integration can be substantially higher than those computed with finite elements, and the resonance frequencies continue to change significantly as the maximum degree of the basis functions is increased, indicating that the method is far from converging. Surface-mesh integration most likely fails in these situations not because of loss of numerical precision, but because it uses global, polynomial displacement functions to approximate mode shapes. For both objects with small features and large holes, even the lowest normal modes involve displacements that vary significantly over small fractions of the object's total length, and these displacements cannot be captured, even approximately, by low-order polynomials. As noted by Yoneda \citet{xyzr}, this is a feature of the polynomial basis employed in the Visscher method and its variants, and not of SMI in particular. Our method will converge to the correct spectrum by extending the basis functions to higher degree, however, floating point precision causes numerical instability well before surface-mesh integration and finite element agree.

\section{\label{sec:4} Conclusion}

We have developed and validated a method to rapidly, repeatedly calculate normal mode spectra for irregularly shaped objects. By performing numerical integration over the object's surface, approximated by a triangular mesh, we extend the variational method developed by \cite{visscher_rus} for objects with simple geometric shapes to objects with irregular geometry. In addition to being implemented entirely with open-source software, surface-mesh integration allows resonance spectra to be computed orders of magnitude faster than finite element methods. For both simple and irregular objects, surface-mesh integration finds resonance spectra within $\sim 10^{-4}$ of both Visscher and finite element methods for the first 100 resonances, which is well within the uncertainty of an RUS experiment. This allows fits of elastic moduli to RUS spectra to be performed on as-grown crystals, eliminating the laborious and sometimes impossible procedure of preparing well-aligned rectangular prisms.

\section{\label{sec:5} Acknowledgements}

The authors acknowledge support for this work and writing the manuscript from the Office of Basic Energy Sciences of the United States Department of Energy under award no. DE-SC0020143. The authors acknowledge support from the Cornell Center for Materials Research with funding from the Materials Research Science and Engineering Centers program of the National Science Foundation (cooperative agreement no. DMR-1719875). The authors acknowledge Teresa Porri and funding from NIH S10OD012287 for use of the ZEISS/Xradia Versa 520 X-ray Microscope to collect CT scans of samples. The authors thank Albert Migliori for helpful discussions. 

\bibliographystyle{unsrtnat}
\bibliography{stokes_ms_literature}

\begin{thebibliography}{29}
\providecommand{\natexlab}[1]{#1}
\providecommand{\url}[1]{\texttt{#1}}
\expandafter\ifx\csname urlstyle\endcsname\relax
  \providecommand{\doi}[1]{doi: #1}\else
  \providecommand{\doi}{doi: \begingroup \urlstyle{rm}\Url}\fi

\bibitem[Ghosh et~al.(2021)Ghosh, Shekhter, Jerzembeck, Kikugawa, Sokolov,
  Brando, Mackenzie, Hicks, and Ramshaw]{sr2ruo4_sc}
Sayak Ghosh, Arkady Shekhter, F.~Jerzembeck, N.~Kikugawa, Dmitry~A. Sokolov,
  Manuel Brando, A.~P. Mackenzie, Clifford~W. Hicks, and B.~J. Ramshaw.
\newblock Thermodynamic evidence for a two-component superconducting order
  parameter in {Sr$_2$RuO$_4$}.
\newblock \emph{Nature Physics}, 17:\penalty0 199, 2021.

\bibitem[Shekhter et~al.(2013)Shekhter, Ramshaw, Liang, Hardy, Bonn, Balakirev,
  McDonald, Betts, Riggs, and Migliori]{ybco_pseudogap}
Arkady Shekhter, B.~J. Ramshaw, Ruixing Liang, W.~N. Hardy, D.~A. Bonn,
  Fedor~F. Balakirev, Ross~D. McDonald, Jon~B. Betts, Scott~C. Riggs, and
  Albert Migliori.
\newblock Bounding the pseudogap with a line of phase transitions in
  {YBa$_2$Cu$_3$O$_{6+\delta}$}.
\newblock \emph{Nature}, 498:\penalty0 75, 2013.

\bibitem[Ramshaw et~al.(2015)Ramshaw, Shekhter, McDonald, Betts, Mitchell,
  Tobash, Mielke, Bauer, and Migliori]{pucoga5_valence}
B.~J. Ramshaw, Arkady Shekhter, Ross~D. McDonald, Jon~B. Betts, J.~N. Mitchell,
  P.~H. Tobash, C.~H. Mielke, E.~D. Bauer, and Albert Migliori.
\newblock Avoided valence transition in a plutonium superconductor.
\newblock \emph{Proceedings of the National Academy of Sciences}, 112\penalty0
  (11):\penalty0 3285--3289, 2015.

\bibitem[Theuss et~al.(2022)Theuss, Ghosh, Chen, Tchernyshyov, Nakatsuji, and
  Ramshaw]{mn3x_strong}
Florian Theuss, Sayak Ghosh, Taishi Chen, Oleg Tchernyshyov, Satoru Nakatsuji,
  and BJ~Ramshaw.
\newblock Strong magnetoelastic coupling in {Mn$_3$X (X= Ge, Sn)}.
\newblock \emph{Physical Review B}, 105\penalty0 (17):\penalty0 174430, 2022.

\bibitem[Tarumi et~al.(2007)Tarumi, Hirao, Ichitsubo, Matsubara, Saida, and
  Kato]{cu_glass}
R.~Tarumi, M.~Hirao, T.~Ichitsubo, E.~Matsubara, J.~Saida, and H.~Kato.
\newblock Low-temperature acoustic properties and quasiharmonic analysis for
  cu-based bulk metallic glasses.
\newblock \emph{Phys. Rev. B}, 76:\penalty0 104206, 2007.

\bibitem[Mejia et~al.(2018)Mejia, Born, Schiemer, Felser, Carpenter, and
  Nicklas]{op_coupling_alloys}
C.~Salazar Mejia, N.-O. Born, J.~A. Schiemer, C.~Felser, M.~A. Carpenter, and
  M.~Nicklas.
\newblock Strain and order-parameter coupling in ni-mn-ga heusler alloys from
  resonant ultrasound spectroscopy.
\newblock \emph{Phys. Rev. B}, 97:\penalty0 094410, 2018.

\bibitem[Luan et~al.(2009)Luan, andRongying Jin, and
  Madrus]{layered_perovskite}
Yanbing Luan, Veerle~Keppens andRongying Jin, and David Madrus.
\newblock Resonant ultrasound studies of the layered perovskite system
  ca$_2-x$sr$_x$ruo$_4$.
\newblock \emph{Journal of the Acoustical Society of America}, 126:\penalty0
  2949, 2009.

\bibitem[Li et~al.(2015)Li, Koehler, Bredeson, He, Mandrus, and
  Keppens]{doubled_perovskite}
Ling Li, Michael~R. Koehler, Isaac Bredeson, Jian He, David Mandrus, and Veerle
  Keppens.
\newblock Magnetoelastic coupling in a$_2$fereo$_6$ (a = ba and ca) probed by
  elastic constants and magnetostriction measurements.
\newblock \emph{Journal of Applied Physics}, 117:\penalty0 213913, 2015.

\bibitem[Luo et~al.(2018)Luo, Lin, Fobes, Liu, Bauer, Betts, Migliori,
  Thompson, Janoschek, and Maiorov]{skyrmion_mnsi}
Yongkang Luo, Shi-Zeng Lin, D.~M. Fobes, Zhiqi Liu, E.~D. Bauer, J.~B. Betts,
  A.~Migliori, J.~D. Thompson, M.~Janoschek, and B.~Maiorov.
\newblock Anisotropic magnetocrystalline coupling of the skyrmion lattice in
  mnsi.
\newblock \emph{Phys. Rev. B}, 97:\penalty0 104423, 2018.

\bibitem[Luo et~al.(2020)Luo, Lin, Leroux, Wakeham, Fobes, Bauer, Betts,
  D.Thompson, Migliori, Janoschek, and Maiorov]{skyrmion_creep}
Yongkang Luo, Shi-Zeng Lin, Maxime Leroux, Nicholas Wakeham, David~M. Fobes,
  Eric~D. Bauer, Jonathan~B. Betts, Joe D.Thompson, Albert Migliori, Marc
  Janoschek, and Boris Maiorov.
\newblock Skyrmion lettice creep at ultra-low current densities.
\newblock \emph{Communication Materials}, 1, 2020.

\bibitem[Petit et~al.(2005)Petit, Duquennoy, Ouaftouh, Deneuville, Jenot, and
  Ourak]{rus_bearings}
S.~Petit, M.~Duquennoy, M.~Ouaftouh, F.~Deneuville, F.~Jenot, and M.~Ourak.
\newblock Nondestructive testing of aeronautic bearing ceramic balls by
  resonant ultrasound spectroscopy.
\newblock \emph{AIP Conference Proceedings}, 760:\penalty0 1251, 2005.

\bibitem[McGuigana et~al.(2021)McGuigana, Arguelles, Obaton, Donmez, Riviere,
  and Shokouhi]{rus_complex}
Samantha McGuigana, Andrea~P. Arguelles, Anne-Francoise Obaton, Alkan~M.
  Donmez, Jacques Riviere, and Parisa Shokouhi.
\newblock Resonant ultrasound spectroscopy for quality control of geometrically
  complex additively manufactured components.
\newblock \emph{Additive Manufacturing}, 39:\penalty0 101808, 2021.

\bibitem[Migliori and Darling(1995)]{rus_testing}
Albert Migliori and Timothy~W. Darling.
\newblock Resonant ultrasound spectroscopy for materials studies and
  non-destructive testing.
\newblock \emph{Ultrasonics International Conference Proceedings}, 1995.

\bibitem[Farrar et~al.(1999)Farrar, Darling, Migliori, and Baker]{rus_bridge}
C.~R. Farrar, T.~Darling, A.~Migliori, and W.~Baker.
\newblock Microwave interferometers for non-contact vibration measurements on
  large structures.
\newblock \emph{Mechanical Systems and Signal Processing}, 13\penalty0
  (2):\penalty0 241, 1999.

\bibitem[Balakirev et~al.(2019)Balakirev, Ennaceur, Migliori, Maiorov, and
  Migliori]{rus_toolbox}
Fedor~F. Balakirev, Susan~M. Ennaceur, Robert~J. Migliori, Boris Maiorov, and
  Albert Migliori.
\newblock Resonant ultrasound spectroscopy: The essential toolbox.
\newblock \emph{Rev. Sci. Instrum.}, 90:\penalty0 121401, 2019.

\bibitem[Migliori and Maynard(2005)]{modern_rus}
Albert Migliori and J.D. Maynard.
\newblock Implementation of a modern resonant ultrasound spectroscopy system
  for the measurement of the elastic moduli of small solid specimens.
\newblock \emph{Rev. Sci. Instrum.}, 76:\penalty0 121301, 2005.

\bibitem[Visscher et~al.(1991)Visscher, Migliori, Bell, and
  Reinert]{visscher_rus}
William~M. Visscher, Albert Migliori, Thomas~M. Bell, and Robert~A. Reinert.
\newblock On the normal modes of free vibration of inhomogeneous and
  anisotropic elastic objects.
\newblock \emph{The Journal of the Acoustical Society of America}, 90:\penalty0
  2154, 1991.

\bibitem[Liu and Maynard(2012)]{fem_arbitrary_rus}
Guoxing Liu and J.~D. Maynard.
\newblock Measuring elastic constants of arbitrarily shaped samples using
  resonant ultrasound.
\newblock \emph{The Journal of the Acoustical Society of America},
  131:\penalty0 2068, 2012.

\bibitem[Plesek et~al.(2004)Plesek, Kolman, and Landa]{fem_fpi_rus}
Jiri Plesek, Radek Kolman, and Michal Landa.
\newblock Using finite element method for the determination of elastic moduli
  by resonant ultrasound spectroscopy.
\newblock \emph{The Journal of the Acoustical Society of America},
  116:\penalty0 282, 2004.

\bibitem[Wang et~al.(2021)Wang, Fan, Shen, Wang, Laugier, and
  Niu]{fem_dif_evolution_rus}
Rui Wang, Fan Fan, Fei Shen, Yue Wang, Pascal Laugier, and Haijun Niu.
\newblock Application of differential evolution on elasticity measurement of
  low quality factor materials using fem-based resonant ultrasound
  spectroscopy.
\newblock \emph{Journal of the Mechanical Behavior of Biomedical Materials},
  124:\penalty0 2068, 2021.

\bibitem[Wang et~al.(2019)Wang, Fan, Zhang, Shen, Laugier, and
  Niu]{fem_muct_rus}
Rui Wang, Fan Fan, Qiang Zhang, Fei Shen, Pascal Laugier, and Haijun Niu.
\newblock Fem-based resonant ultrasound spectroscopy method for measurement of
  the elastic properties of irregular solid specimens.
\newblock In \emph{IEEE International Ultrasonics Symposium}, 2019.

\bibitem[Wu(2000)]{bem_wu}
T.~W. Wu.
\newblock \emph{Boundary Element Acoustics: Fundamentals and Computer Codes}.
\newblock W. I. T. Press, Southampton, 2000.

\bibitem[Demarest(1971)]{demarest_rus}
Harold~H. Demarest.
\newblock Cube-resonance method to determine the elastic constants of solids.
\newblock \emph{The Journal of the Acoustical Society of America}, 49:\penalty0
  768, 1971.

\bibitem[Rathod and Rao(1996)]{tetrahedron_int}
H.~T. Rathod and H.~S.~Govinda Rao.
\newblock Integration of polynomials oer an arbitrary tetrahedron in euclidian
  three-dimensional space.
\newblock \emph{Computers and Structures}, 59\penalty0 (1):\penalty0 55--65,
  1996.

\bibitem[Jaśkowiec and Sukumar(2020)]{tetrahedron_quad}
Jan Jaśkowiec and N.~Sukumar.
\newblock High-order cubature rules for tetrahedra.
\newblock \emph{International Journal for Numerical Methods in Engineering},
  121\penalty0 (11):\penalty0 2418--2436, 2020.

\bibitem[Beer et~al.(2008)Beer, Smith, and Duenser]{bem_theory}
Gernot Beer, Ian Smith, and Christian Duenser.
\newblock \emph{The Boundary Element Method with Programming}.
\newblock Springer, Vienna, 2008.

\bibitem[Kirkup(1998)]{bem_acoustics}
Stephen Kirkup.
\newblock \emph{The Boundary Element Method in Acoustics}.
\newblock Integrated Sound Software, 1998.

\bibitem[Albuquerque et~al.(2003)Albuquerque, Sollero, and
  Fedelinski]{bem_anisotropy}
EL~Albuquerque, P~Sollero, and P~Fedelinski.
\newblock Free vibration analysis of anisotropic material structures using the
  boundary element method.
\newblock \emph{Engineering Analysis with Boundary Elements}, 27:\penalty0
  977--985, 2003.

\bibitem[Yoneda(2000)]{xyzr}
Akira Yoneda.
\newblock The xyzr algorithm for eigenvibration problem of bored and liminated
  objects.
\newblock \emph{Journal of Sound and Vibration}, 236:\penalty0 431--441, 2000.

\end{thebibliography}

\end{document}